# Test item response time and the response likelihood

Srdjan Verbić[1] & Boris Tomić
*Institute for Education Quality and Evaluation*

*Test takers do not give equally reliable responses. They take different responding strategies and they do not make the same effort to solve the problem and answer the question correctly. The consequences of differential test takers' behavior are numerous: the test item parameters could be biased, there might emerge differential item functioning for certain subgroups, estimation of test taker's ability might have greater error, etc. All the consequences are becoming more prominent at low-stakes tests where test takers' motivation is additionally decreased. We had analyzed a computer based test in Physics and tried to find and describe relationship between the item response time and the item response likelihood. We have found that magnitude of such relationship depends on the item difficulty parameter. We have also noticed that boys, who respond faster, in average, give responses with greater likelihood than the boys who respond slower. The same trend was not detected for girls.*

**Introduction**

Computer based tests (CBT) enable us to routinely record times of students' responses to each particular test item. This data could be important additional source of information both on the test taker's abilities and the test characteristics. The response time is considered to be important component of ability and drew attention to the need to develop testing models that would include test takers' response time, that is, the amount of time a test taker spends in reading and answering question. Many authors of the previous studies considering test item response time (Hornke, 2005; Schnipke & Scrams, 1997; Scrams & Schnipke, 1999; Wang & Hanson, 2005) have pointed that further research in this area can improve the efficiency of cognitive tests by offering additional information about the impact of the question on a test taker.

*Low-stakes tests and test takers' motivation*

In exploratory large-scale assessment students are often asked to take achievement tests for which they receive neither grades nor academic credit. At low-stakes tests students may little care if their test scores do not represent their true levels of proficiency because they would not receive sanctions for poor test performance, and because strong test performance would not help them get something they want (Wise & Kong, 2005). In such cases, the motivation levels of test takers become a matter of concern to test givers because a lack of examinee effort represents a direct threat to the reliability of the test data. If a test taker does not try hard, obtained results would underestimate the test taker's true level of profi-ciency. Furthermore, if estimate student's proficiency on the basis of inadequate effort presented in a sample of test data, estimates of item characteristics will be biased (Wise, 2006).

---

[1] Corresponding author: Srdjan Verbić, Institute for Education Quality and Evaluation, Fabrisova 10, 11000 Belgrade, Serbia; e-mail: sverbic@ceo.edu.rs.

Although the intent of assessments of student achievement is to measure what is called "maximum performance" they can do so only if students attempt to do their best while taking the test. If a student is not motivated to put forth a reasonable effort, the results may clearly underestimate his or her maximal proficiency and so lead to invalid interpretations of the obtained scores and a distortion of the results (Abdelfattah, 2007).

Furthermore, it is more likely that examinees will try harder on certain kinds of items. For example, items with graphs are more likely to receive test taker effort (Wise, Kong, & Pastor, 2007), while items that require open responses are less likely to engender optimal motivation. Thus, test takers appear to be discriminating in how they approach exerting effort on low-stakes tests. However, further research is needed to explore which testing characteristics are associated with increased motivation and which are associated with lower motivation (Lau & Pastor, 2007).

*Response behaviors and scoring schemas*

The test takers, generally, behave in two basic ways: solving the problem or guessing the answer. When we use low-stakes tests, low motivation of examinees causes behaviors different from problem-solving or rational guessing like rapid-guessing or rapid-omitting. Rapid-guessing is common behavior at speeded high-stakes tests (Schnipke & Scrams, 1997) as well as at majority of low-stakes tests (Wise & Kong, 2005). Main reason for rapid-guessing at high-stakes tests is the lack of time, while lack of test taker's motivation is the main reasons at low-stakes tests.

When engaged in solution behavior, test takers read each item carefully and fully consider the solution. Consequently, response times arising from solution behavior will likely depend on item length, difficulty, and other characteristics, as well as on person-specific variables. Accuracy will depend jointly on test taker's ability and item difficulty and other item characteristics (Schnipke & Scrams, 1997). If test takers do not find solution in reasonably long time, they will generally behave in more economic way and try to guess the answer or to omit the item. Rational guessing or omitting also takes time and it could depend on the item characteristics. Such a behavior is much different from rapid-responding (rapid guessing or rapid omitting) where examinees just skim items briefly for keywords, but they do not thoroughly read the items.

At low stakes tests, rapid-guessers are typically unmotivated examinees who answer too quickly, i.e. before they have time to read and fully consider the item. The accuracy of their responses is close to accuracy of random guessing. Depending on instructions given to test takers, omitting an item can be alternative to guessing. Rapid guessing is a common problem with reliability and validity of low-stakes tests (Wise, 2006; Wise & Kong, 2005). Rapid-omitting is just a manifestation of the same phenomenon.

Rapid-responding behavior could be diminished if we instruct students to consider possibility of non-answering because non-answer could bring them more points than the wrong answer. Here we can use Traub's rule (Traub, Hambleton, & Singh, 1969) as an instruction for students and the scoring schema for preliminary results provided in feedback to the students. Although the scoring scheme for Traub's rule is equivalent to formula scoring[2], there is some empirical evidence that Traub's rule yields higher reliabilities and validities than formula scoring. In addition, test takers seem to prefer award for non-answer to penalty for wrong answer. The answer why would two scoring formulas that are strategically equivalent and perfectly correlated result in different psychometric properties lies in the psychological dimension (Budescu & Bar-Hillel, 1993).

---

[2] Formula scoring levies a penalty of $1/(m-1)$ points against each incorrect answer, yielding a final score of $S=n_R-n_W/(m-1)$ where $n_R$ represents number of correct answers, $n_W$ number of wrong answers, and $m$ number of options. This is the rule employed by the Scholastic Aptitude Test (SAT) since 1953, as well as by the GRE subject exams (Budescu & Bar-Hillel, 1993).

A rational test taker, whose goal is to maximize score on the test, answering strategy is either superior or equivalent to omitting. For test takers who are not fully rational, or have goals other than the maximization of expected score, it is very hard to give adequate formula scoring instructions, and even the recommendation to answer under partial knowledge is problematic, though generally beneficial (Budescu & Bar-Hillel, 1993).

Reducing of rapid-responding behavior should give us more realistic item response times which depend on item characteristics. This feature of item responses is very important for reliable parameter estimation. Speededness does not affect ability estimation significantly, but it affects parameter estimation (Oshima, 1994). The same thing we can expect for all kinds of rapid-responding. Furthermore, if we measure typical response times, we could explore reasons and effects of unexpected response times which can be indicative of specific types of aberrant behavior (van der Linden & van Krimpen-Stoop, 2003).

*Modeling item response time*

Item response time (RT), or item latency in some literature, is generally defined as the time elapsed between presenting the question on the computer screen and the response to that question. Many previous studies suggest that item response times are well fitted by the lognormal distributions (Schnipke & Scrams, 1999; Thissen, 1983; van der Linden, 2006). We can normalize response time data with natural logarithmic transformation to create a more normal distribution required by most statistical procedures. The lognormal probability density function (PDF) of item response time ($t$) is given by

$$\text{PDF}_{\text{lognormal}}(t) = \frac{1}{t\sigma\sqrt{2\pi}} \exp\left[-\frac{(\log t - \mu)^2}{2\sigma^2}\right]. \tag{1}$$

Applying the lognormal density to the raw data is equivalent to applying the normal density to the logarithm of the raw data. The parameters of the lognormal density were estimated by taking the mean ($\hat{\mu}$) and standard deviation ($\hat{\sigma}$) of natural logarithm of response times. Depending on the test (used items and the context) and the examinees' characteristics, the resulting density function can be bimodal. Observed response-time distributions can be described as a mixture of response-time distributions of different examinees' behaviors (Schnipke & Scrams, 1997).

*Response likelihood*

Binary response model allows correct or incorrect answers only. If we know the item parameters and test taker's ability, we can estimate probability that the test taker would answer to the item correctly or incorrectly. We expect that more able test taker answers correctly to an easy item as well as we expect that less able examinee answers incorrectly to a hard one. This expectancy can be quantified using likelihood ($L$) of the response to the item:

$$L(\theta) = \pi^s(\theta)(1-\pi(\theta))^{1-s}, \tag{2}$$

where $\pi$ stands for probability that examinee with ability $\theta$ gives correct answer and $s$ is score of the examinee's response to the item which is 1 for correct or 0 for incorrect answer. That way we can calculate how likely is that certain examinee gives one particular answer. Responses with low likelihood are unlikely and we can be suspicious how well they reflect test taker's true ability.

Estimation of π is based on two parameter logistic item response theory (2PL IRT) model. Therefore, we calculate π using formula:

$$\pi(\theta) = \frac{1}{1+e^{-a(\theta-b)}}, \qquad (3)$$

where *a* represents item discrimination and *b* item difficulty parameter obtained from 2PL IRT analysis. We have estimated test taker's ability θ using maximum likelihood algorithm.

*Response time–likelihood relationship*

Results of previous studies show that relationship between response times and different response measures depends heavily on test context and content. Examinees put different effort to high- or low-stakes tests. At low-stakes tests we have significant number of rapid-responders while problem-solvers take time to respond. At high-stakes tests effort plays more important role and ultimately we increase chances to answer correctly if we spend more time on the particular item. Researchers have found that speed and accuracy on complex tasks do not measure the same construct. Items' complexity, i.e. number of steps needed to find correct answer, greatly influences relationship between item response time and probability to answer correctly.

Schnipke & Scrams (2002) concluded that cognitive psychologists have tended to focus on the within-person relationship between response speed and response accuracy (speed-accuracy tradeoff) – if a person responds to an item more quickly, would the person's accuracy tend to decline? On the other hand, psychometric researchers have tended to focus more on the across-person relationship between speed and accuracy – do the most accurate test takers tend to respond slower or faster than their less accurate counterparts? Speed-accuracy relationships has been modeled by making ability partially dependent on the time devoted to the item (Verhelst, Verstralen, & Jansen, 1997). This type of speed-accuracy relationship is modeled explicitly in Thissen's (1983) timed-testing model by the coefficients relating the logarithm of response time to the IRT logit. Thissen explored the relationships on three cognitive tests. Correlation between log RT and IRT logit was positive for all of them. Schnipke & Scrams (1997) applied a version of Thissen's timed-testing model to responses to three skill tests. All tests resulted in very small values of the coefficient of correlation between log RT to IRT logit.

Response accuracy defined as probability that test taker with ability θ gives correct answer does not take into account what is the actual test taker's response. The response likelihood is a measure of reliability test taker's response for a given response model. It is not likely that high-ability student gives incorrect answer to easy, highly discriminative item. Such an occurrence could be explained by lower level of effort that examinee applied to the item solving. On the other hand, if low-ability student gives correct answer to hard, highly discriminative item, that could be a consequence of pure guessing or some other type of aberrant responding behavior. Likelihood function gives us information how likely is that a test taker give one particular response. Likelihood as a measure of the response reliability is based on assumption that all test takers use the same strategy for responding to an item. If test takers use different strategies for the same item, the response time-likelihood relationship would depend greatly on the applied strategies.

*Gender difference in response time and DIF*

Different response strategies can have different characteristic response times. Strategy choice could be related to different groups of examinees. There are only a few studies concerning differential response time for different subgroups. Previous research in this area was focused mainly on differences in rapid-guessing behavior. Schnipke & Scrams explored this problem for speeded high-stakes tests where examinees had to respond quickly because lack of time. Wise and his colleagues explored rapid-guessing at low-stakes tests. In this case examinees responded quickly because of lack of motivation to work hard on some items.

Examinee subgroups may differ in the rates at which they tend to work on a timed high-stakes test (Llabre & Froman, 1987). Oshima (1994) discussed that when test may be differentially speeded for the two groups, differential amounts of strategic rapid-guessing behavior occur at the end of the test, which could potentially lead to differential item functioning (DIF).

It has been reported recently that girls do not respond rapidly at low-stakes test as much as boys do (Wise, Kingsbury, Thomason, & Kong, 2004). Such a behavior is assumed to be one of possible reasons for DIF. Schnipke (1995) studied gender differences relative to effort; she found that rapid-guessing behavior was common among male examinees on an analytical test. In addition, rapid guessing was more common among female examinees on a quantitative test, and equally common on a verbal test. This difference in speed could be the cause of bias of item parameters and estimation of test reliability, as well as cause of DIF.

Results of DeMars & Wise (2007) also suggest that DIF can be caused by differences in examinee effort. When trying to understand why a particular item exhibited DIF, measurement practitioners should therefore consider questions focused both on item content (i.e., what is it about the item's content that could have produced DIF?) and examinee behavior (i.e., why would one group have given less effort to this item than another group?). Thus, the results of this investigation have illustrated that item response time can play an important role in guiding measurement practitioners to better understand the causes of DIF and whether or not item deletion is warranted when DIF occurs.

Assumed difference in characteristic response times for boys and girls could be the cause of differential item functioning. The question is whether this difference is large enough to be detected through the Mantel-Haenszel DIF procedure (Holland & Thayer, 1986). This procedure gives us significance of DIF. All items with $p<0.05$ were flagged as exhibiting DIF favoring one of two groups.

*Research questions*

In this study, we tried to answer four research questions:
1) Can we find and describe relationship between item response time and likelihood of item responses?
2) Does such relationship depend on item difficulty or item discrimination?
3) Is there a difference in characteristic response time and response likelihood for boys and girls?
4) Can we detect differential item functioning using differential response time for boys and girls?

**Method**

*Target group and the instrument*

We have conducted off-line computer based testing in Physics for 352 eight grade students (164 boys and 188 girls) from 18 schools in Serbia. The test contained 32 items dealing with understanding of waves (sound and light) and elements of laboratory measurement (data use, analysis and interpretation). We have used three types of items defined in Moodle: closed (multiple choice), semi-closed (numerical) and open-ended (essay).

This was a trial test, a part of pilot study for introduction of large-scale computerized assessment in Serbia realized in 2006-2008. Metric characteristics of majority of used items were not previously determined. All students had the same set of items arranged in several different ways because of test security reasons. It was pure power-test where all students had enough time to respond all the items.

The software environment for test delivery was Moodle, open-source software for producing internet-based courses (Dougiamas, 2001). Moodle has a module named "Quiz" which is designed for classroom assessment and does not fulfill all our requirements for exploratory large-scale assessment. Therefore, we have developed another version of this module: Ttest (Tomić, Lacković, & Verbić, 2008). It is based on Quiz but provides us not only with students' scores but all data including time required that student answer each particular question.

*Procedure*

All test takers were asked to think about all the questions and to give their best answer although the results will be used for the exploration purposes only. They were also told that they will receive their score afterward. Before the test students filled up the questionnaire in the same Web form as the test, prepared so students could practice answering different question types before the test.

Since we were interested in time that students are spending while they think about questions and answer them, we instructed students to consider possibility of non-answering because non-answer could bring them more points than wrong answer according to modified Traub's rule that we used here. Students were awarded with 0.25 points for all non-answered items, does not matter whether it was multiple-choice, numerical or constructed response item. That way we tried to prevent frequent occurrence of rapid-responding. Students were asked to work through the test by their own pace. This scoring schema is used for the feedback to students only. For the research purposes, we used IRT analysis.

*Measuring response time*

Each question was presented on a separate Web page, which allowed measuring the time spent on a particular question – a measurement technique employed by Bergstrom et al. (1994). Although by default the items were presented in numerical order, examinees could answer the items in any order by using navigation tools. They could omit items (i.e., see an item but not answer it and proceed to another), and they could go back to previously viewed items and change their answers. The test was administered in this way so that it would be as similar as possible to the paper-and-pencil mode of test delivery. The recorded response time for an item was the total time spent on the item during all attempts as it was proposed by Schnipke & Scrams (1997). Response times were acquired with 1 second accuracy. For the purpose of this study the time of the first approach and the accumulated time for all approaches to a question were recorded for each test taker and each item.

*Data analysis*

The response data (answers and times) were analyzed using R: a language for statistical computing (R Development Core Team, 2007) with ltm package for Latent Variable Modelling and Item Response Theory Analyses (Rizopoulos, 2006), irtoys interface package (Partchev, 2008), and IRT command language (Hanson, 2002).

## Results

*Rapid responding behavior*

We have identified rapid-responding behavior for fairly small proportion of students: 0.3% of all responses took less than 5 seconds. This result shows that we had much lower level of rapid responding than previously reported for low-stakes tests: 5% (Wise & DeMars, 2006), 5-15% (Yang, 2007). This result is probably the most obvious consequence of the instruction that discourage rapid and random responding. Distribution of item response time logarithms is given at Figure 1.

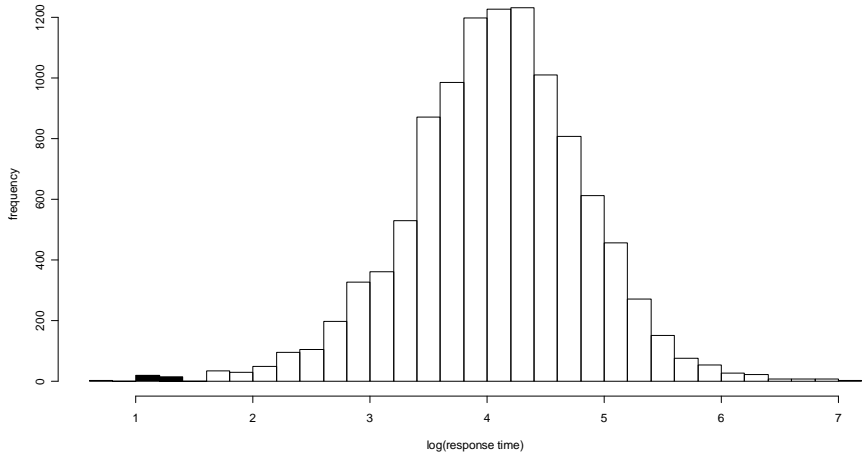

FIGURE 1 Histogram of all item response time logarithms. Filled bars on the left side of the histogram represents unusually quick responses (*t*<5 s).

*Item response likelihood*

Response likelihoods were calculated for all items and all students. It is hard to see any meaningful relationship between log RT and likelihood for a particular student. If we look at the relationship for particular items, we could see that correlation between log RT and response likelihood has different values for different items. We have found that correlation between log RT and likelihood is negative for almost all used items. There was no evidence that correlation coefficient depends on item type or the item discrimination parameter. However, we have noticed that magnitude of such relationship is depending on the item difficulty parameter (*b*). This relationship is presented at Figure 2. Pearson's correlation between log RT and response likelihood for all items, excluding inadequate items (five items with difficulty parameter greater than 3 or less than -3), correlation is *R*=-0.70. Two items with lowest values of correlation between log RT and response likelihood (less than -0.3) are one numerical item where fast responders frequently give the same wrong answer (#4) and one open-ended item requiring brief explanation (#32).

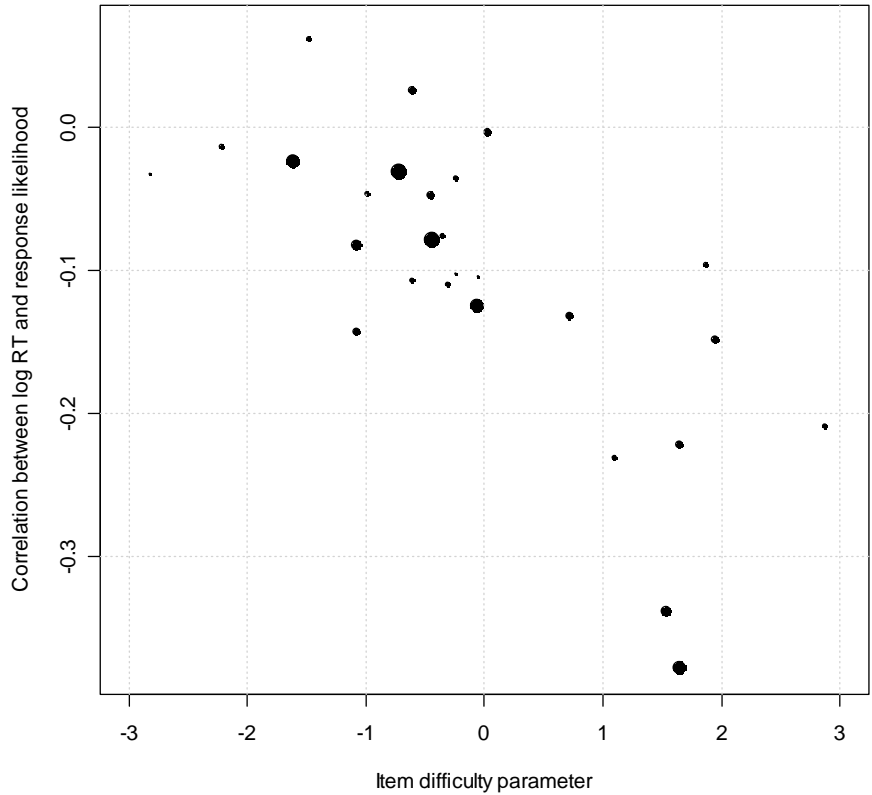

FIGURE 2 Relationship between item difficulty parameter and correlation of log RT and response likelihood for test items with difficulty between -3 and 3. Diameter of circles on the graph is proportional to item discrimination parameter (*a*).

*Differential response time for boys and girls*

There is significant difference between characteristic response times of boys and girls on this test. Girls were answering correctly to an item, in average, 7 seconds later. Their incorrect answers and non-answers also took longer time: 7 and 14 seconds respectively. These results are similar to those previously reported for 4th grade students in subject Nature & Society by Verbić & Tomić (2008).

|  | correct answer | incorrect answer | omitting |
|---|---|---|---|
| boys | 3.99 | 4.06 | 3.84 |
| girls | 4.12 | 4.18 | 4.11 |

TABLE 1: Average item response time logarithm for boys and girls (Physics, Grade 8)

We can also notice that boys gave correct answers faster than girls to 30 out of 32 items. They responded faster to 22/32 items by giving incorrect answer and to 27/32 by omitting. Generally, boys responded faster to 31 out of 32 items.

Difference in response times of boys and girls did not affect their achievement. We can not tell that boys perform better than girls or vice versa, but we certainly can say that boys respond faster.

In order to eliminate difference between characteristic times for different items, we have scaled all log RT's so that scaled log RT has average value equal to zero for all items. Distribution of scaled log RT's is given on Figure 3 (left). Difference between mean transformed time values for boys and girls is very significant ($p<0.005$).

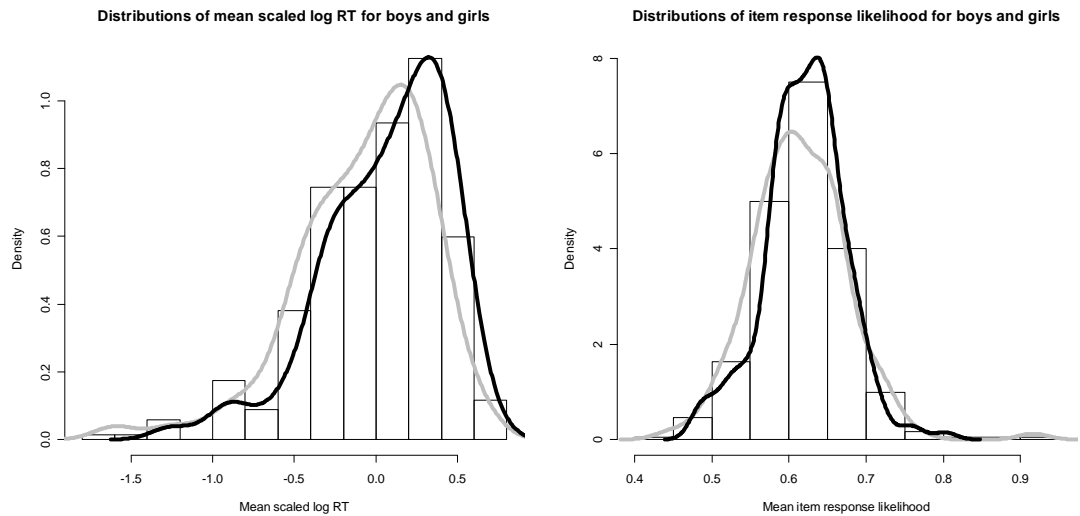

FIGURE 3 Histograms of mean scaled log response times (left) and likelihood of item responses (right) for all examinees. The curves represent distribution densities for male (gray) and female (black) examinees.

*Differential response likelihood for boys and girls*

Mean value of item response likelihood is slightly greater for girls (0.621) than for boys (0.614). This difference is not statistically significant. We have noticed that there is significant difference between the distributions' width. Variance of mean item response likelihood is 1.46 times greater for girls than for boys ($p<0.02$).

Lack of significant difference between mean response likelihoods for boys and girls does not imply that there is no difference locally, at some fixed response time. As a measure of each particular student characteristic response time, we can use mean log RT, or $e^{\text{mean}(\log RT)}$ if we want RT expressed in seconds. This value is better measure of the sample than mean value because of asymmetry of lognormal distribution of response times. Figure 4 shows mean response likelihood and characteristic response time[3] given as $e^{\text{mean}(\log RT)}$ for all students. Careful look at the graph reveals that, in spite of great likelihood variance, gray circles (boys) have trend of going down with increase of time. We can not see the same trend for black circles (girls).

---

[3] Since $e^{\text{mean}(\log RT)} \approx \text{median}(RT)$ if RT follows lognormal distribution, we can think of characteristic response times as of median values of RT.

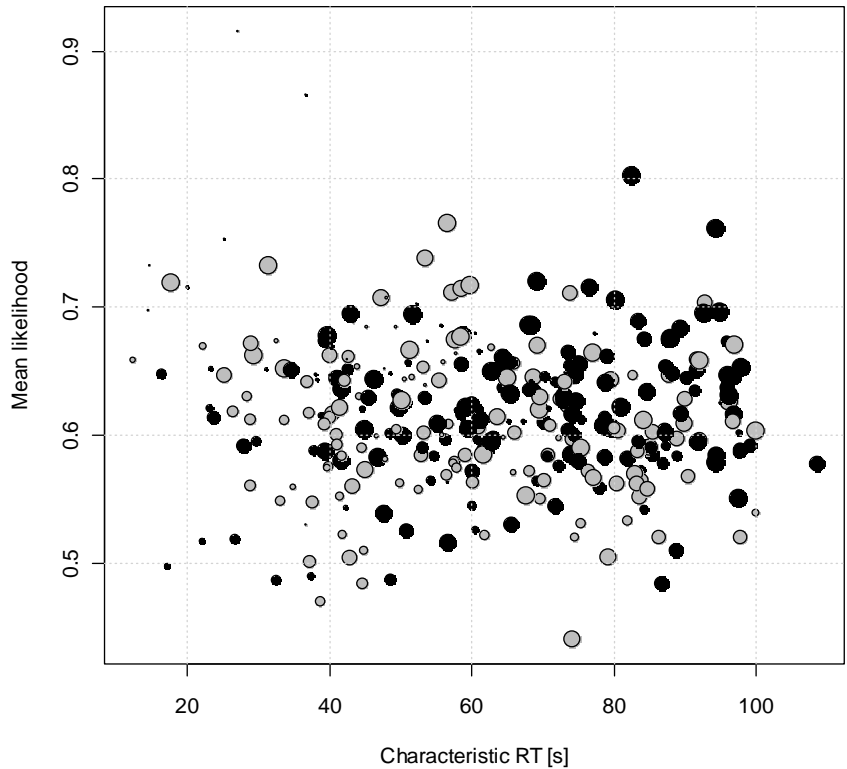

FIGURE 4 Relationship between characteristic RT and mean response likelihood for all test takers. Gray circles denote boys and the black girls. Area of the circles on the graph is proportional to the achievement percentile at the test.

When we calculate Pearson's correlation coefficient between characteristic RT and mean response likelihood for all students, we obtain weak likelihood decreasing trend: $R_{all}=-0.08\pm0.05$ with error cited at 68% confidence level. If we apply linear regression here, that would imply that average response likelihood would decrease by 0.005 for 20 seconds difference in characteristic response time. This trend is significantly different for boys and girls alone: $R_{boys}=-0.22\pm0.08$ and $R_{girls}=+0.04\pm0.08$. Fast boys' responses are, in average, more likely than for slow ones. Girls' responses seem to be equally likely in spite of different response times.

*Response time and differential item functioning*

There are items with great difference in log RT for boys and girls, but we can not see how that influences difference in their achievement. We have found DIF for four items but we could not anticipate which item would be DIF flagged on the basis of differential response times only.

**Discussion**

Variances of item response time within- or across-students are too big to enable us to predict accuracy or likelihood of the item response for a particular student. If we observe relationship between log RT and response likelihood on the item level, correlation between these two variables appears to be distinctive characteristic of an item. For the most of items in our test, the correlation coefficient was negative which means that test takers who respond faster give responses with greater likelihood. This item property becomes more ap-parent for harder items. All results obtained here should be scrutinized in circumstances where students take low-stakes, power test without time-limit pressure.

Significant occurrence of rapid-responders certainly would mask this relationship because rapid-responders give nearly random responds which, in average, have very low likelihood. That is the reason why it is needed to encourage students to think about the answer at their own pace and evaluate whether it is better to guess or to omit the item.

We also believe that single item response time and response likelihood depends much on the item quality, i.e. discrimination parameter. Low average likelihood tells us more about the item quality than about the test takers. Similarly, long response times are rather consequence of bad and non-informative than good and hard item.

The item response time is, probably, the consequence of test taker's confidence in submitted answer. We think that tests items like those we analyzed here, provoke students to give their responses faster (1) if they are sure about the answer (even if they are wrong) or (2) if they realize that they do not know the answer and that they should respond fast and move to another item. This behavior is more common among boys than among girls. We suppose that described behavior could be much different for high-stakes tests or test with items requiring more complex problem solving.

We have demonstrated also that there was a difference in responding behavior between boys and girls at the test we have analyzed. Difference in mean likelihood between boys and girls is not significant, but we can show that shapes of the likelihood distribution differ for these two groups. Trend of likelihood decreasing with amount of time needed to respond is more evident within boys that within girls. Likelihood of item response for girls does not seem to depend significantly on the response time.

Generally, boys respond to an item faster than girls. Girls take 10-15% more time to respond to each item. This occurrence does not influence their test scores. Probability of differential item functioning does not seem to be affected by difference in the response time at power, low-stakes tests.

**Conclusions**

Previous studies mainly explored relationships between test taker's response times and probability of giving correct answer. We have tried to find out whether there is a relationship between the response time and the response likelihood. Key difference between these two approaches is that the later approach gives estimated probability not for expected but for the given answer. This is particularly important when low-stakes tests when students do not work at "maximum performance" level. We have found that magnitude of such relationship is depending on the item difficulty parameter.

Evidences that variances of the response time and the response likelihood are too big to enable us to make useful predictions imply that we need larger samples and strictly controlled item properties in order to reliably describe relationship between the item response time and the item likelihood. Response behaviors certainly depend on many factors like type of the test, type of the items, whether there is time limit or not, motivation of test takers, instructions given to the test takers, etc. Here we had this relationship described in the case of low-stakes computer-based test in Physics intended to examine conceptual knowledge without unnecessary calculations. Generally, we can see the difference in the response likelihood between fast and slow responders. This property is more emphasized for harder items and male group of test takers.

Difference in distributions of response time and response likelihood between boys and girls is evident. This difference in response behavior could be a consequence of different level of risk-aversion and selection of response strategies. We think that described behavior could be much different for item with complex calculations, heavy reading load, etc. Future research in this area would have to cover wide ranges of item statistical and cognitive properties.

Relationship between the response time and the response likelihood could provide us with valuable additional information about test properties and test takers' behavior. Development of efficient item response time model and detailed analysis of response time in test results would increase efficiency of computer-base testing and open new problems for research and further applications.

**References**

Abdelfattah, F. A. (2007). *Response latency effects on classical and item response theory parameters using different scoring procedures.* Ohio University.
Bergstrom, B., Gershon, R. C., & Lunz, M. E. (1994). *Computerized Adaptive Testing Exploring Examinee Response Time Using Hierarchical Linear Modeling.* Paper presented at the Annual Meeting of the National Council on Measurement in Education.
Budescu, D., & Bar-Hillel, M. (1993). To Guess or Not to Guess: A Decision-Theoretic View of Formula Scoring. *Journal of Educational Measurement, 30*(4), 277-291.
DeMars, C. E., & Wise, S. L. (2007). *Can Differential Rapid-Guessing Behavior Lead to Differential Item Functioning?* Paper presented at the Annual meeting of the American Educational Research Association.
Dougiamas, M. (2001). Moodle: open-source software for producing internet-based courses. [Computer software]. Available at http://www.moodle.org
Hanson, B. A. (2002). IRT Command Language. [Computer software]. Available at http://www.b-a-h.com/software/irt/icl
Holland, P. W., & Thayer, D. T. (1986). *Differential Item Performance and the Mantel-Haenszel Procedure*. Paper presented at the Annual Meeting of the American Educational Research Association.
Hornke, L. F. (2005). Response time in computer-aided testing: a" Verbal Memory" test for routes and maps. *Psychology Science, 47*(2), 280.
Lau, A. R., & Pastor, D. A. (2007). *A Hierarchical Linear Model of Variability in Test Motivation Across Students and Within Students Across Tests*. Paper presented at the Meeting of the Northeastern Educational Research Association.
Llabre, M. M., & Froman, T. W. (1987). Allocation of Time to Test Items: a Study of Ethnic Differences. *Journal of Experimental Education, 55*(3), 137-140.
Oshima, T. C. (1994). The Effect of Speededness on Parameter Estimation in Item Response Theory. *Journal of Educational Measurement, 31*(3), 200-219.
Partchev, I. (2008). Simple interface to the estimation and plotting of IRT models. [Computer software]. Available at http://cran.r-project.org/web/packages/irtoys/index.html
R Development Core Team. (2007). R: A language and environment for statistical computing. Vienna, Austria: R Foundation for Statistical Computing.
Rizopoulos, D. (2006). ltm: An R Package for Latent Variable Modeling and Item Response Theory Analyses. *Journal of Statistical Software, 17*(5), 1–25.
Schnipke, D. L. (1995). *Assessing Speededness in Computer-Based Tests Using Item Response Times*. Paper presented at the Annual meeting of NCME.
Schnipke, D. L., & Scrams, D. J. (1997). Modeling Item Response Times With a Two-State Mixture Model: A New Method of Measuring Speededness. *Journal of Educational Measurement, 34*(3), 213-232.
Schnipke, D. L., & Scrams, D. J. (2002). Exploring issues of examinee behavior: Insights gained from response-time analyses. In M. T. P. Craig N. Mills, John J. Fremer, William C. Ward (Ed.), *Computer-based testing: Building the foundation for future assessments* (pp. 237–266): Lawrence Erlbaum Associates.
Scrams, D. J., & Schnipke, D. L. (1999). *Making use of response times in standardized tests: are accuracy and speed measuring the same thing?* Newtown, PA: Law School Admission Council.
Thissen, D. (1983). Timed testing: An approach using item response theory. In D. J. Weiss (Ed.), *New horizons in testing: Latent trait test theory and computerized adaptive testing* (pp. 179-203). New York: Academic Press.
Tomić, B., Lacković, I., & Verbić, S. (2008). Ttest module. [Computer software]. Available at http://sepp.ceo.edu.rs/moodle/mod/resource/view.php?id=25

Traub, R. E., Hambleton, R. K., & Singh, B. (1969). Effects of promised reward and threatened penalty on performance of a multiple-choice vocabulary test. *Educational and Psychological Measurement, 29*(4), 847-861.

van der Linden, W. J., & van Krimpen-Stoop, E. (2003). Using response times to detect aberrant responses in computerized adaptive testing. *Psychometrika, 68*(2), 251-265.

Verbić, S., & Tomić, B. (2008). *Applicability of computer based assessment in primary school*. Paper presented at the XIV annual meeting "Empirical research in psychology", Belgrade, Serbia.

Verhelst, N. D., Verstralen, H., & Jansen, M. G. H. (1997). A logistic model for time-limit tests. In W. J. van der Linden & R. K. Hambleton (Eds.), *Handbook of modern item response theory* (pp. 169-185).

Wang, T., & Hanson, B. A. (2005). Development and Calibration of an Item Response Model That Incorporates Response Time. *Applied Psychological Measurement, 29*(5), 323.

Wise, S. L. (2006). An Investigation of the Differential Effort Received by Items on a Low-Stakes Computer-Based Test. *Applied Measurement in Education, 19*(2), 95-114.

Wise, S. L., & DeMars, C. E. (2006). An Application of Item Response Time: The Effort-Moderated IRT Model. *Journal of Educational Measurement, 43*(1), 19-38.

Wise, S. L., Kingsbury, G. G., Thomason, J., & Kong, X. (2004). *An investigation of motivation filtering in a statewide achievement testing program*. Paper presented at the Annual meeting of the National Council on Measurement in Education, San Diego, CA.

Wise, S. L., & Kong, X. (2005). Response Time Effort: A New Measure of Examinee Motivation in Computer-Based Tests. *Applied Measurement in Education, 18*(2), 163-183.

Wise, S. L., Kong, X. J., & Pastor, D. A. (2007). *Understanding correlates of rapid-guessing behavior in low-stakes testing: Implications for test development and measurement practice*. Paper presented at the Annual meeting of the National Council on Measurement in Education.

Yang, X. (2007). Methods of Identifying Individual Guessers from Item Response Data. *Educational and Psychological Measurement, 67*(5).

APPENDIX 1

| Item number | Item type | Item discrimination parameter, $a$ | Correlation between log RT and the response likelihood | Difference between average log RT for boys and girls | DIF $p$-value for boys and girls |
|---|---|---|---|---|---|
| #1 | MC | 0.82 | -0.15 | -0.12 | 0.65 |
| #2 | MC | 0.74 | -0.02 | -0.03 | 0.54 |
| #3 | MC | 0.41 | -0.10 | -0.09 | 0.48 |
| #4 | NR | 1.58 | -0.38 | -0.16 | 0.41 |
| #5 | CR | 0.65 | -0.23 | -0.22 | 0.33 |
| #6 | MC | 0.72 | -0.16 | -0.15 | 0.07 |
| #7 | MC | 0.33 | -0.05 | **-0.34** | **0.01** |
| #8 | MC | 0.69 | -0.04 | -0.21 | 0.17 |
| #9 | MC | 0.45 | -0.11 | -0.22 | **0.01** |
| #10 | CR | 0.72 | -0.11 | -0.09 | 0.84 |
| #11 | MC | 0.34 | -0.10 | -0.19 | 0.50 |
| #12 | MC | 0.37 | -0.03 | -0.14 | 0.72 |
| #13 | MC | 0.61 | -0.07 | -0.10 | 0.49 |
| #14 | MC | 0.91 | -0.22 | **-0.31** | 0.49 |
| #15 | MC | 0.82 | -0.13 | -0.19 | **0.02** |
| #16 | MC | 0.57 | -0.05 | -0.09 | 0.82 |
| #17 | MC | 0.45 | -0.09 | **-0.28** | 0.95 |
| #18 | MC | 0.78 | -0.14 | -0.21 | 0.86 |
| #19 | CR | 0.56 | -0.21 | -0.02 | 0.46 |
| #20 | MC | 0.33 | 0.07 | -0.01 | 0.46 |
| #21 | MC | 1.78 | -0.02 | -0.04 | 0.91 |
| #22 | MC | 0.91 | 0.03 | -0.04 | 0.52 |
| #23 | MC | 1.26 | -0.08 | -0.09 | 0.28 |
| #24 | NR | 0.82 | 0.00 | -0.06 | 0.36 |
| #25 | NR | 0.59 | 0.06 | **0.01** | 0.86 |
| #26 | NR | 0.43 | 0.02 | -0.10 | 0.80 |
| #27 | NR | 1.35 | -0.03 | -0.11 | 0.75 |
| #28 | NR | 1.36 | -0.12 | -0.15 | 0.53 |
| #29 | NR | 0.59 | -0.18 | -0.20 | 0.65 |
| #30 | NR | 1.89 | -0.08 | -0.13 | 0.84 |
| #31 | NR | 0.82 | -0.04 | -0.06 | 0.77 |
| #32 | CR | 1.11 | -0.34 | -0.18 | 0.60 |

TABLE 2 Item parameters related to the response behavior. Item types are labeled with MC for multiple-choice, CR for constructed response and NR for numerical question. There are three items where differential item functioning is significant (, #7, #9, and #15), there are three items where boys' response times were by more than 30% shorter than girls', and only one item (#25) where girls responded to the item (insignificantly) faster.